\documentclass{iopart}
\usepackage{mathptmx}
\usepackage{bm,amsfonts}
\makeatletter
\@ifundefined{pdfoutput} %\% Definitely not using pdftex.
{ %\% Standard TeX
\usepackage[dvips]{graphicx,color}
}
{ %\% Running pdftex.
\ifnum\pdfoutput=0\relax% Are we outputting pdf?
\usepackage[dvips]{graphicx,color}
\fi
\ifnum\pdfoutput=1\relax% Are we outputting pdf?
\usepackage[pdftex]{graphicx,color}
\fi
}
\makeatother
\newcommand{\beq}{\begin{equation}}
\newcommand{\eeq}{\end{equation}}
\newcommand{\beqa}{\begin{eqnarray}}
\newcommand{\eeqa}{\end{eqnarray}}
\newcommand{\nn}{\nonumber \\}
\newcommand {\np}[1]{{\mbox{\textrm{:}\,}{#1}{\,\textrm{:}}} }
\def \e {\mathrm{e}}

\def \la {\langle}
\def \ra {\rangle}
\def \s {\sigma}

\def \B {{\mathcal B}}

\def \I {{\mathbb I}}

\def \Z {{\mathbb Z}}

\def \qh {\mathrm{qh}}

\begin{document}
%\begin{frontmatter}
%%%%%%%%%%%%%%%%%%%%%%%%%%%%%%%%%%%%%%%%%%%%%%%%%%%%%%%%%%%%
\title[Ultimate Ising braid-group generators ]{Ultimate braid-group generators for coordinate exchanges of Ising 
anyons from the multi-anyon Pfaffian wave functions}
%%%%%%%%%%%%%%%%%%%%%%%%%%%%%%%%%%%%%%%%%%%%%%%%%%%%%%%%%%%%
\author{Lachezar S. Georgiev}
 \address{Institut f\"ur Mathematische Physik, Technische Universit\"at Braunschweig,
Mendelssohnstr. 3, 38106 Braunschweig, Germany,}
\address{Institute for Nuclear Research and Nuclear Energy, Bulgarian Academy of Sciences, 
 Tsarigradsko Chaussee 72,  1784 Sofia, Bulgaria}
\begin{abstract}
We give a rigorous and self-consistent derivation of the elementary braid matrices representing the exchanges of
adjacent Ising anyons in the two inequivalent representations of the Pfaffian quantum Hall states with even and odd
number of Majorana fermions. 
To this end we use the distinct  operator product expansions of the chiral spin fields in the Neveu--Schwarz and 
Ramond sectors of  the two-dimensional Ising conformal field theory. We find recursive relations for the
generators of the irreducible representations of the braid group $\B_{2n+2}$ in terms of those for $\B_{2n}$, 
as well as explicit formulas for almost all braid matrices for exchanges of Ising anyons.
Finally we prove that the braid-group representations obtained from the multi-anyon Pfaffian wave functions
are completely equivalent to the spinor representations of SO$(2n+2)$ and give the equivalence matrices explicitly.
This actually proves that the correlation functions of $2n$ chiral Ising spin fields $\s$ do indeed realize one of the 
two inequivalent spinor representations of the rotation group SO$(2n)$ as conjectured by Nayak and Wilczek.
\end{abstract}

%\maketitle
%%%%%%%%%%%%%%%%%%%%%%%%%%%%%%%%%
\section{Introduction}
%%%%%%%%%%%%%%%%%%%%%%%%%%%%%%%%%%
One fascinating application \cite{sarma-freedman-nayak,freedman-nayak-walker} of the anticipated non-Abelian 
statistics of the chiral spin 
fields in the critical two-dimensional Ising model has established a remarkable connection between the 
rich-of-exact-results area of the two-dimensional rational conformal field theories (CFT) and 
the new and promising field of topological  quantum computation \cite{sarma-RMP}. 
The localized non-Abelian Ising anyons, which are believed to be realized in the fractional quantum Hall state  at filling
 factor $\nu=5/2$, that is most likely described by the  Moore--Read (Pfaffian) CFT \cite{mr}, turned out to be a useful tool 
for topologically protected quantum information processing 
\cite{sarma-freedman-nayak,freedman-nayak-walker,TQC-NPB,sarma-RMP}. 
Protection against noise and decoherence is obtained by encoding quantum information in robust topological 
characteristics of the strongly correlated electron system, such as quasiparticle's fusion channels,  while quantum gates 
are implemented by topologically non-trivial operations 
\cite{sarma-freedman-nayak,freedman-nayak-walker,TQC-NPB,clifford}, 
such as braidings of non-Abelian anyons.

Nayak and Wilczek argued in an insightful paper that the Pfaffian 
wave functions with $2n$ Ising anyons at fixed positions belong to a $2^{n-1}$ dimensional spinor
representation of the rotation group SO$(2n)$, see Sect. 9 in Ref.~\cite{nayak-wilczek}. 
However, as explained in Sect.~\ref{sec:NW} below, the arguments they presented in support of this claim were 
incomplete and partly misleading. To our knowledge, this claim therefore has been, up to now, only a conjecture. 
In the present paper we provide a complete proof of this conjecture meeting the requirements of rigor of mathematics.
This proof contains three steps:
\begin{enumerate}
\item{
Derive the elementary generators, $B_j^{(4,\pm)}$, $j=1,2,3$, of the braid group $\B_4$, in the 
two inequivalent irreducible representations with positive (``$+$'' in the superscript) and negative 
(``$-$'' in the superscript) fermion parity, directly from the 4-quasihole Pfaffian wave functions.}
\item{
Construct recursively the generators  $B_j^{(2n+2,\pm)}$, $1\leq j \leq 2n+1$,  of the irreducible representations of 
the braid group $\B_{2n+2}$ in terms of the generators, $B_j^{(2n,\pm)}$, $1\leq j \leq 2n-1$, of $\B_{2n}$.}
\item{
Find explicitly the equivalence matrices mapping the obtained generators of the representations of the 
braid group  $\B_{2n+2}$ to those  generating the representations of Nayak--Wilczek with the corresponding parity. }
\end{enumerate}
Contrary to the common believe, the first step has not been accomplished by NW in 
Ref.~\cite{nayak-wilczek}---they have only computed the first line of the $2\times 2$  matrix representing the exchange 
of the anyons with coordinates $\eta_1$ and $\eta_3$,
which at best could be used to determine the generators of the positive-parity representation of $\B_4$,
though some ambiguities had to be resolved, see Sect.~\ref{sec:NW} below.
Later the braid matrix $R_{23}$, as well as the other two generators $R_{12}$ and  $R_{34}$, for the exchanges of 
4 Ising anyons have been unambiguously  derived in  Ref.~\cite{TQC-NPB} by careful analytic continuation of 
the 4-quasiholes Pfaffian wave functions in the representation corresponding to even number of fermion fields in the 
CFT correlation function. 

However, the generators of the negative-parity representation of $\B_4$ have never been derived before, in 
the wave-function approach, because the 4-anyon Pfaffian wave functions with odd number of Majorana fermions 
have been unknown. In this paper we give the first, to our knowledge,  derivation of the generators of the 
negative-parity representation of $\B_4$ directly from the Pfaffian wave functions, without computing them explicitly, 
by using instead the short-distance operator product expansions.

Furthermore, the generators of the two inequivalent representations of the braid group $\B_{2n+2}$, spanned 
by the Pfaffian wave functions realized as CFT correlation functions with $2n+2$ Ising anyons at fixed positions
and even/odd number of Majorana fermions, are not easy to obtain. They cannot be derived in analogy
with the generators of $\B_4$ because the Pfaffian wave functions with $2n+2$ anyons are not known explicitly. 
Fortunately, it is possible to find recursion relations between the generators of $\B_{2n+2}$ and those of $\B_{2n}$,
by using the fusion rules of the Ising anyons. However, when we fuse two Ising anyons, the result could be either $\I$
or a Majorana fermion $\psi$ so that in the first case the fusion process maps a representation of $\B_{2n+2}$ with  
given fermion parity into a representation of $\B_{2n}$ with the same parity, while in the second case the fusion 
process  switches to the opposite parity.  This subtlety  not only mixes the representations but also requires that 
we know both representations of  $\B_{2n}$ in order to construct inductively from them any of the representations of  
$\B_{2n+2}$.

Meanwhile, the 4-quasihole results of NW have been reproduced in Ref.~\cite{slingerland-bais} using 
the universal  $R$ matrix in the quantum-group approach for the Ising model. It is worth stressing that the 
arguments of \cite{slingerland-bais} do not prove the NW conjecture because the representation of the braid 
group $\B_4$ in \cite{slingerland-bais} is \textit{defined by the authors} in such a way to reproduce the results 
of  NW as can be seen from Sects. 5.2 and 5.3 in \cite{slingerland-bais}, however, there is no proof that 
it is the same as the braid-group representation obtained by analytic continuation of the multi-anyon Pfaffian 
wave functions as obtained e.g. in \cite{TQC-NPB}.

In addition, the 4-anyon braid matrices have been convincingly derived in Ref.~\cite{ivanov} for the case of the 
$p$-wave superconductor, which is known to be related to the Pfaffian state \cite{read-green,stern-oppen-mariani}.
While the question of the basis orthonormality in the $p$-wave superconductor has not been addressed 
in \cite{ivanov},  it has been answered in  \cite{stern-oppen-mariani}. 
The analysis of the $p$-wave superconductor is more conclusive about the braid generators, 
however, the connection to the Pfaffian FQH state is more elaborate, because only the large-distance behavior of 
the weak-pairing phase is related to the MR state \cite{read-green}.

The above mentioned confirmations would have been very nice if we had an independent proof of the NW 
conjecture,  however, they are  still not sufficient to prove this conjecture, because they are either defined on 
purpose to reproduce the NW results or are indirectly related to the many-body states of the electron system 
and the correspondence depends on many assumptions. 
Therefore, it would be useful to have an independent, self-consistent  and rigorous derivation of the braid matrices 
directly from the Pfaffian wave functions representing the states containing multiple Ising anyons, which are actually 
used to define the qubits in TQC \cite{sarma-freedman-nayak}. 

One more reason for this necessity is that the eventual experiments with the real quantum Hall systems 
would test the properties of the strongly correlated electron state that are encoded into the corresponding 
many-body wave function.
Recall that the gauge-invariant quantity in the adiabatic transport exchanging Ising quasiholes is the 
product of the explicit monodromy (which can be computed in the CFT or quantum group approach) and the 
geometrical Berry phase \cite{gurarie-nayak,simon-LL-mixing,read-viscosity} which is present only in the 
wave-function approach. This is a subtle point because whether the adiabatic transport of Ising anyons, along 
complete loops around each  other, is indeed  realized by spinor representations of the rotation group depends on
the Berry connection of the trial wave functions. It has been argued that the actual holonomy, which is 
the physically observable quantity that we intend to use for topological quantum computation, is indeed equal to 
the monodromy of the Pfaffian wave functions because the Berry connection is trivial  
\cite{gurarie-nayak,stern-oppen-mariani,simon-LL-mixing,read-viscosity}, i.e., the  only contribution comes 
from the ubiquitous Gaussian factor that is typical for charged particles in magnetic field and this geometrical 
contribution is simply the Aharonov--Bohm phase. Given that the multi-anyon trial wave functions are holomorphic,
that would certainly be true if they could be proven to be orthonormal \cite{simon-LL-mixing,read-viscosity}.
 While the first attempts \cite{gurarie-nayak} succeed in generalizing the analogy of (the overlap screening of) the 
Coulomb plasma, at least to the Pfaffian state with two anyons, a recent argument about the four-anyon case
\cite{read-viscosity}, which could be generalized to more anyons, seems to provide convincing evidence that 
the multi-anyon wave functions obtained in an appropriate CFT basis are indeed orthonormal. 
Therefore, it is now rather plausible 
that the holonomy of the multi-anyon Pfaffian wave functions is precisely given by the monodromy which could 
be obtained by analytic continuation, and this is what we shall use in this paper.
Notice also that the Landau level mixing, which has important consequences  for the physics of the quantum Hall 
state at filling factor $\nu=5/2$, would certainly  modify\cite{simon-LL-mixing} the exchange properties of the Ising 
anyons derived by the monodromies and this effect can only be analyzed in the wave-function approach.

The details in the explicit representation of the braid generators and the differences between their distinct realizations
become more important when we try to implement various quantum gates and to estimate the computational power of the 
Ising-anyon TQC \cite{TQC-NPB}. For example, it was possible to construct the CNOT gate \cite{TQC-NPB} in terms of 7 
elementary braidings, however this construction  could not be generalized for systems with more anyons, i.e.,
 the CNOT could not be embedded into systems with more qubits. The precise braid-generators analysis 
 is crucial for answering such questions as whether it is possible to implement the entire Clifford-gate group 
purely by braiding or not, see Ref. \cite{clifford} for the answer.

\textit{Outline of the paper:}
In this paper we give  a rigorous and unified derivation of the braid matrices,
representing the exchanges of Ising anyons, and a proof of the NW conjecture based on the 
wave-function approach, and this result is independent of the orthonormality of the CFT blocks used as a basis.
Combined  with the orthonormality results obtained in Ref.~\cite{read-viscosity} this implies that the adiabatic transport 
of Ising anyons could indeed be used for topological quantum computation as proposed in 
\cite{nayak-wilczek,sarma-freedman-nayak,stern-oppen-mariani,read-viscosity}. 
In Sect.~\ref{sec:R-sector} we review the subtle issue of the chiral fermion parity and its spontaneous breaking 
in the doubly degenerate Ramond sector of the Ising model due to the presence of the Majorana fermion zero mode, 
following  Refs.~\cite{fst,5-2} and introducing their notation. This is necessary for the formulation of the 
short-distance operator product expansions of the spin fields $\s_\pm$ in the Ramond sector, derived in \cite{fst}, 
which is the main tool for
the computation of the non-diagonal braid matrices $B_{2}^{(4,\pm)}$ in Sects.~\ref{sec:B_23+} and \ref{sec:B_23-}.
In Sect.~\ref{sec:wave-funcs} we recall the definitions of the 4-quasihole Pfaffian wave functions as correlation 
functions \cite{mr,nayak-wilczek} in the $\widehat{u(1)}\times$Ising  CFT containing four $\s$ fields and even number 
of Majorana fermions and introduce the notation for the computational states, as well as the encoding 
of quantum information, following Refs.~\cite{TQC-NPB,clifford}. 
In Sect.~\ref{sec:NW}  we formulate the NW conjecture, which we intend to prove in this paper, and in 
Sect.~\ref{sec:results}  we announce the precise statement about the braid generators that we obtain in 
the Pfaffian wave-function representations of the braid group $\B_4$ with positive and negative parity, 
which is the first step towards the proof.
Then in Sect.~\ref{sec:diag} we derive the diagonal braid matrices $B_1^{(4,\pm)}$ and $B_3^{(4,\pm)}$
by using the short-distance operator product expansion (OPE) in the  Neveu--Schwarz (NS) sector of the Ising model,
 representing the elementary 4-anyon exchanges in the positive-parity (Sect.~\ref{sec:diag+}) and negative-parity 
(Sect.~\ref{sec:diag-}) wave-function representations. Next, in Sect.~\ref{sec:non-diag} we derive the non-diagonal 
braid matrices $B_2^{(4,\pm)}$, in the wave-function representations with positive parity (Sect.~\ref{sec:B_23+}) 
and negative parity (Sect.~\ref{sec:B_23-}), by using the short-distance OPE, this time in the Ramond (R) sector 
of the Ising model, which is reviewed in Sect.~\ref{sec:OPE-R} following Ref.~\cite{fst}.
The braid generators $B_j^{(4,+)}$, $j=1,2,3$,  obtained in the representation of $\B_4$ with positive fermion parity,
 coincide completely with those derived in Ref.~\cite{TQC-NPB} for the case when the 
homotopic  condition $| \eta_{12} \eta_{34} / \eta_{13} \eta_{24} |< 1$ of  Ref.~\cite{TQC-NPB} is fulfilled.
Notice that $B_j^{(4,+)}$ obtained here are explicitly different from those of NW.
 Furthermore, the braid generators $B_j^{(4,-)}$, $j=1,2,3$,  for the representation of $\B_4$ with negative 
fermion parity, are important new results which have not been obtained before from the 4-quasihole Pfaffian wave 
functions  because these wave functions were unknown in the negative-parity representations.
In the Proposition 1 in Sect.~\ref{sec:induct}  we derive new recursive relations
for the projected $(2n+2)$-anyon exchange generators  $B_j^{(2n+2,\pm)}$,  $j=1,\ldots, 2n+1$, in terms of 
those for $2n$ anyons, $B_j^{(2n,\pm)}$,  $j=1,\ldots, 2n-1$,  and give maximally 
explicit new formulas for the (projected) braid matrices in the Corollary.
Finally we prove in the Proposition 2 in Sect.~\ref{sec:proof} that the braid-group representations derived from 
the multi-anyon Pfaffian wave functions
are equivalent to those derived in the SO$(2n+2)$ spinor  approach and give the explicit equivalence matrices
which are also new results.
This equivalence makes it completely legitimate to interpret the abstract parity in the spinor representations of 
SO$(2n+2)$ as the physical fermion parity in the Ising model.
%%%%%%%%%%%%%%%%%%%%%%%%%%%%%%%%%%%%%%%%%%%%%%%%%%%%%%%%%%%%
\subsection{Double degeneracy of the R-sector and spontaneous breaking of fermion parity}
\label{sec:R-sector}
%%%%%%%%%%%%%%%%%%%%%%%%%%%%%%%%%%%%%%%%%%%%%%%%%%%%%%%%%%%%
The Ramond sector (or twisted sector) of the Ising model is defined as the superselection sector in which the 
Majorana fermion field 
has  periodic boundary conditions on the cylinder and antiperiodic on the conformal plane \cite{fst,CFT-book}. 
Because of the periodic boundary conditions of the Majorana fermion in the R-sector, whose Laurent mode expansion 
in the complex plane
\[
\psi(z)=\sum_{n\in \Z} \psi_n z^{n-1/2}, \quad \{ \psi_n, \psi_m\}=\delta_{n+m,0}
\]
 implies the presence of a fermionic zero mode $\psi_0$,  the ground state in this sector must necessarily 
be doubly degenerate if the chiral fermion parity is conserved. 
 Indeed, the Majorana zero  mode $ \psi_0=\psi_0^\dagger$ , $\left(\psi_0\right)^2= (1/2)\I $
and the chiral fermion parity operator  $\gamma$, satisfying 
$\gamma \psi_0 +\psi_0 \gamma =0$ and $\gamma^2=\I$, 
form a two-dimensional  Clifford algebra  whose lowest dimensional representation is two-dimensional and can 
be expressed in terms of the Pauli matrices \cite{5-2}.
Choosing a $\gamma$-diagonal basis,  the two chiral spin fields of CFT dimension $1/16$
(with positive and negative fermionic parity)
 intertwining between the vacuum and the R-sector's lowest weight state can  be written as \cite{fst}
\beq \label{sigma_pm}
|\pm\ra = \s_{\pm} (0) |0\ra, \quad \gamma \s_{\pm} \gamma = \pm \s_{\pm}.
\eeq
The conservation of the fermion parity implies that the two fields  $\s_\pm$ in Eq.~(\ref{sigma_pm}) must obey 
Abelian fusion rules
\beq \label{fusion-NS}
\s_+ \times \s_+ =\I, \quad \s_+ \times \s_- =\psi, \quad  \s_- \times \s_- =\I , 
\quad \psi \times \s_\pm = \s_\mp.
\eeq
On the other hand, modular invariance requires a single lowest-weight state  \cite{5-2},
like in the case of the Gliozzi--Scherk--Olive projection in string theory,  which is 
conventionally chosen as
\[
\s = \frac{ \s_{+} +\s_{-}}{\sqrt{2}} \quad \Longrightarrow \quad \s \times \s = \I +\psi
\]
and consequently obeys non-Abelian statistics. This projection leads to spontaneous breaking of the chiral 
fermion parity, see Ref.~\cite{5-2} for a more detailed explanation. Despite 
the  seemingly unphysical nature of the chiral spin fields $\s_{\pm}$ with definite fermion parity they appear to be 
very convenient for enumerating different  computational states, 
for labeling the fusion paths in the corresponding Bratteli diagrams \cite{sarma-RMP,clifford},
as well as for the identification of the spinor parity in the representations of SO$(2n+2)$ with the fermion parity in
the Ising model.
%%%%%%%%%%%%%%%%%%%%%%%%%%%%%%%%%%%%%%%%%%%%%%%%%%%%%%%%%%%%
\subsection{Wave functions for 4 Ising anyons in the positive-parity representation}
\label{sec:wave-funcs}
%%%%%%%%%%%%%%%%%%%%%%%%%%%%%%%%%%%%%%%%%%%%%%%%%%%%%%%%%%%%
The wave function  for the Pfaffian fractional quantum Hall state with even number  $N$ of holes (or electrons) at positions
$z_1, \ldots, z_N$ and four quasiholes at positions $\eta_1,\ldots, \eta_4$, can be realized as a correlation function,
in the $\widehat{u(1)}\times \textrm{Ising}$ CFT \cite{mr,nayak-wilczek,TQC-NPB},
\beq\label{4-qh}
\Psi_{4\qh}(\eta_1,\eta_2,\eta_3,\eta_4; \{z_i\})=
\la \psi_{\mathrm{qh}}(\eta_1)\psi_{\mathrm{qh}}(\eta_2)\psi_{\mathrm{qh}}(\eta_3)
\psi_{\mathrm{qh}}(\eta_4)\prod_{i=1}^N
\psi_{\mathrm{hole}}(z_i) \ra
\eeq
 of the field operators corresponding to creation of holes and quasiholes
\beq \label{fields}
\psi_{\mathrm{hole}}(z)= \psi(z) \, \np{\e^{i \sqrt{2}\phi(z)}  } \quad
\mathrm{and}\quad
\psi_{\mathrm{qh}}(\eta)= \s(\eta)\, \np{\e^{i\frac{1}{2\sqrt{2}} \phi(\eta)}},
\eeq
respectively, where $\sigma(\eta)$ is the chiral spin field in the Ising model
of dimension $1/16$ and  $\psi(z)$ is the right-moving Majorana fermion
 in the chiral Ising model.
 It can be expressed in more explicit form in terms of the Pfaffian wave functions \cite{TQC-NPB}, however, 
because they will not be needed in our fusion-rules  approach we will skip them.
We shall only use the notation of  Ref.~\cite{nayak-wilczek,TQC-NPB} for the two linear independent  4-quasiholes states 
\beq\label{01}
\Psi_{4\mathrm{qh}}^{(0)}\equiv |0\ra_+ \sim \la \s_+\s_+\s_+\s_+\ra  , \quad 
\Psi_{4\mathrm{qh}}^{(1)}\equiv |1\ra_+  \sim \la \s_+\s_-\s_+\s_-\ra
\eeq
 and will call them the computational basis in the positive-parity representation (representation parity is denoted by 
the subscript ``$+$'').
 In the next section we will give more detailed expressions for the computational basis states (\ref{01}) 
as well as an explanation of the sign "$\sim$".
We can now argue  why it is important to derive the braid generators directly from the wave functions 
representing FQH states with 4 Ising anyons.
It might indeed be possible to derive the braid-group generators from the universal $R$-matrix for the quantum 
group corresponding to the Pfaffian FQH state \cite{slingerland-bais}. However, it can  be shown that the Pfaffian 
wave functions (\ref{01}) with 4 Ising anyons, defined in Ref.~\cite{TQC-NPB}, are different from the correlation 
functions of four  $\s$ fields (which are zero unless $ e_1 e_2 =e_3 e_4=\kappa$ and then, see Eq.~(6.43) in 
Ref.~\cite{fst})
\beq\label{4-pt}
\la \s_{e_1}(\eta_1)\sigma_{e_2}(\eta_2)\sigma_{e_3}(\eta_3)\sigma_{e_4}(\eta_4) \ra = \frac{1}{\sqrt{2}}
\left(\frac{\eta_{13}\eta_{24}}{\eta_{12}\eta_{14}\eta_{23}\eta_{34}}\right)^{1/8}
\sqrt{1+\kappa \sqrt{x}},
\eeq
so that the former functions would have different analytic properties from the latter because of the presence of 
the Majorana fermions.
Because the 4-point correlation functions (\ref{4-pt}) depend only on the product of the signs of the fields comprising the 
first and the second pair we can encode information in the topological charge $\kappa=e_1 e_2$ of the first pair 
and then the topological charge of the second pair is fixed to be the same. It is also obvious from Eq.~(\ref{4-pt})  
that the order of signs in the two pairs is irrelevant, i.e., $\s_+\s_-\sim \s_-\s_+$ because $e_1e_2 = e_2 e_1=\kappa$,  
so that we can always choose the sign of the first $\s$ field in each pair to be `$+$'.
%%%%%%%%%%%%%%%%%%%%%%%%%%%%%%%%%%%%%%%%%%%%%%%%%%%%%%%%%%%%
\subsection{The Nayak--Wilczek conjecture}
\label{sec:NW}
%%%%%%%%%%%%%%%%%%%%%%%%%%%%%%%%%%%%%%%%%%%%%%%%%%%%%%%%%%%%
Nayak and Wilczek conjectured \cite{nayak-wilczek} that the elementary
matrices representing the exchanges of $2n$ Ising quasiparticles in the
Pfaffian fractional quantum Hall state can be interpreted as $\pi/2$ rotations from SO$(2n)$,
i.e., they can be expressed in terms of the gamma matrices
$\gamma_i^{(n)}$,  $1\leq i \leq 2n$, which satisfy the anticommutation relations
of the Clifford algebra
\beq \label{cr}
\left\{\gamma_i^{(n)},\gamma_j^{(n)} \right\} = 2\delta_{ij}, \quad 1\leq i,j \leq 2n.
\eeq
In more detail,
the elementary operations for the exchange of the $i$-th and $(i+1)$-th quasiparticles
could be expressed, in an appropriate basis of $2n$-quasiholes Pfaffian wave functions 
\cite{nayak-wilczek,ivanov},  as
\beq \label{R}
R_{j}^{(n)} = \e^{i\frac{\pi}{4}}
\exp\left(-\frac{\pi}{4} \gamma_j^{(n)}\gamma_{j+1}^{(n)}\right) \equiv
\frac{\e^{i\frac{\pi}{4}}}{\sqrt{2}}
\left(\I -\gamma_j^{(n)}\gamma_{j+1}^{(n)}\right),
\eeq
where $1\leq j\leq 2n-1$ and  the second  equality follows from the fact that $(\gamma_j\gamma_{j+1})^2=-\I$
due to the anticommutation relations (\ref{cr}).

The $2n$  matrices $\gamma^{(n)}_i$ have dimension $2^n \times 2^n$ and
can be defined explicitly as follows
\cite{wilczek-zee,nayak-wilczek,slingerland-bais}
\beqa \label{gamma-ex}
\gamma_{1}^{(n)} &=& \sigma_1 \otimes \sigma_3\otimes \cdots \otimes\sigma_3   \nn
\gamma_{2}^{(n)} &=& \sigma_2 \otimes \sigma_3\otimes \cdots \otimes\sigma_3  \nn
&\vdots& \nn
\gamma_{2j-1}^{(n)} &=& \underbrace{\I_{2} \otimes \cdots \otimes \I_{2}}_{j-1}\otimes
\ \sigma_1 \otimes \underbrace{\sigma_3\otimes\cdots \otimes\sigma_3 }_{n-j}  \nn
\gamma_{2j}^{(n)} &=&  \underbrace{\I_{2} \otimes  \cdots \otimes \I_{2}}_{j-1}\otimes
\ \sigma_2 \otimes \underbrace{\sigma_3\otimes\cdots\otimes\sigma_3}_{n-j}  \nn
&\vdots& \nn
\gamma_{2n-1}^{(n)} &=&  \I_{2^{n-1}}\otimes \sigma_1 \nn
\gamma_{2n}^{(n)} &=&  \I_{2^{n-1}} \otimes\sigma_2.
\eeqa
The ``gamma-five'' matrix $\gamma_F^{(n)}$,  defined by
\[
\gamma_F^{(n)} =(-i)^n \gamma_1^{(n)} \cdots  \gamma_{2n}^{(n)},
\]
commutes with all matrices (\ref{R}) and therefore $R_j^{(n)}$ cannot change the 
$\gamma_F^{(n)}$ eigenvalues $\pm 1$, which means that the representation (\ref{R}) is reducible
and the two irreducible components, corresponding to  eigenvalues $\pm 1$, can be obtained
by projecting with the two projectors 
\beq \label{P_pm}
P_{\pm}^{(n)} =\frac{\I_{2^n} \pm\gamma_F^{(n)} }{2}, \ \mathrm{i.e.,}
\ \left( P_{\pm}^{(n)}\right)^2= P_{\pm}^{(n)} =\left( P_{\pm}^{(n)}\right)^\dagger .
\eeq
In other words, the generators of the two irreducible spinor representations of the braid group
$\B_{2n}$ can be obtained by simply projecting ($1\leq j\leq 2n-1$)
\beq \label{R_pm}
R_{j}^{(n,\pm) } =
P_{\pm}^{(n)} R_{j}^{(n) } P_{\pm}^{(n)}   
= \frac{\e^{i\frac{\pi}{4}}}{\sqrt{2}}\left(\I -\gamma_j^{(n)}\gamma_{j+1}^{(n)}\right)
P_{\pm}^{(n)} .
\eeq
Eq. (\ref{R_pm}) is what we call the NW conjecture in this paper because the braid generators 
(\ref{R_pm}) have not  been derived in Ref.~\cite{nayak-wilczek} from the multi-anyon Pfaffian wave functions. 
Instead, NW say on page 546 in \cite{nayak-wilczek}:  
``We will verify this assertion in the four-quasihole case with our explicit wave functions  \ldots and give an 
argument in favor of its validity in the $2n$-quasihole case''.  
To this end they first verify the statement for 4 quasiholes by computing the first row of the $2\times 2$ 
braid matrix representing 
the exchange $R_{13}$ of the anyons  with coordinates   $\eta_1$ and  $\eta_3$ and then compute $R_{23}$ from that. 
However,  this kind of derivation of the elementary braid matrix  $R_{23}$ is ambiguous because 
it is based on the result for the exchange of anyons 1 and 3 and  a braid relation, such as 
$R_{23}=R_{12} R_{13}R_{12}^{-1} $ or $R'_{23}=R_{12}^{-1} R_{13}R_{12} $, and the two results are physically 
different. 
Formally, this ambiguity   appears because the exchange $\eta_1 \leftrightarrow \eta_3$ depends on the homotopy 
class of  the exchange with respect to the second anyon (with coordinate $\eta_2$) and because of the emerging 
sign ambiguities which have not been fixed in a physical way.
Next, the generalization argument they mention at the end of Sect. 9 in \cite{nayak-wilczek} is as follows: 
``... imagine bringing 4 quasiholes 
close together ...; the braiding is governed by the OPE and therefore is generated be the transformations  we found 
above in the 4-quasihole case''. 
This argument is misleading because NW found the 4-quasihole  braid generators only for even fermion 
parity of ``the rest'', while their generalization argument assumes that they could use them also in the negative-parity 
case which is wrong. The point is that when we separate 4 quasiholes 
the rest $2n-4$ quasiholes could have both positive and negative total fermion parity (with equal occurrence in 
the computational basis). For example, consider the 6-anyon computational states given in Eq.~(38) in
Ref.~\cite{TQC-NPB}:  following the NW ``generalization argument'',  let us concentrate on the first four 
quasiholes, corresponding to coordinates $\eta_1,\ldots,\eta_4$; the rest of the quasiholes, i.e., the two quasiholes 
with coordinates $\eta_5$ and $\eta_6$ in this case,  have positive total fermion parity in the computational states
denoted by $|00\ra$ and $|10\ra$ in Eq. (38) in \cite{TQC-NPB}, while in the states $|01\ra$ and $|11\ra$ it is negative. 
In the first case one could eventually use the four-quasihole results of 
Ref.~\cite{nayak-wilczek}, while for negative parity one needs an inequivalent set of generators which have been
 missing in \cite{nayak-wilczek}. 
Actually the braid generators in the negative-parity representations of
the braid group $\B_4$ have never been known before, because the explicit four-anyon Pfaffian wave functions in 
the negative-parity  representation have been unknown. Hence one cannot derive recursively the generators of 
$\B_{2n}$ only from those of the positive-parity representation of $\B_4$, so that NW have drawn an inference from 
slight or insufficient evidence. In this paper we are filling this logical gap thus turning the insightful 
NW conjecture into a mathematical theorem. 
%%%%%%%%%%%%%%%%%%%%%%%%%%%%%%%%%%%%%%%%%%%%%%%%%%%%%%%%%%%%
\subsection{Braid matrices for exchanges of 4 Ising anyons: statement of the result}
\label{sec:results}
%%%%%%%%%%%%%%%%%%%%%%%%%%%%%%%%%%%%%%%%%%%%%%%%%%%%%%%%%%%%
One of the important results in this paper is the unified derivation of the braid matrices, $B_1^{(4,\pm)}$, $B_2^{(4,\pm)}$ 
and $B_3^{(4,\pm)}$, in the two inequivalent  representations of the braid group $\B_4$  corresponding to positive 
and negative fermion parity, directly from the four-anyon Pfaffian wave functions. 
The superscript ``4'' in the notation for the braid generators expresses that these 
are generators of representations of $\B_4$, while the sign  ``$\pm$'' denotes the fermion parity of the corresponding
 representation. 
In Sections~\ref{sec:diag} and \ref{sec:non-diag}  below we will show that the result for the generators of the 
positive-parity representation of $\B_4$
\beq \label{R4p}
B_{1}^{(4,+)}=\left[ \matrix{1 & 0 \cr 0 & i}\right],
\quad
B_{2}^{(4,+)}=
\frac{\e^{i\frac{\pi}{4}} }{\sqrt{2}}
\left[ \matrix{\ \    1 & -i \cr -i & \ \ 1}\right],\quad
B_{3}^{(4,+)}=\left[ \matrix{1 & 0 \cr 0 & i}\right]
\eeq
obtained here by the fusion-rules approach completely coincides with the result of Ref.~\cite{TQC-PRL} obtained by 
the analytic continuation of the 4-quasiholes wave function when $|\eta_{12}\eta_{34}/\eta_{13}\eta_{24}|<1$. 

The braid generators $B_j^{(4,+)}$ obtained here are explicitly different from those of NW, albeit the latter
 can be obtained by an equivalence transformation generated by $(B_3^{(4,+)})^2$, see Sect.~\ref{sec:induct} 
below. However,
 this monodromy transformation makes an observable difference for the physical state the topological quantum 
computer, which has important physical consequences, e.g., it implements the Pauli $Z$ gate, and therefore 
has to be controlled experimentally.

In addition we shall explicitly derive the generators of the 4-quasihole Pfaffian representation of the  braid group  $\B_4$ 
with negative fermion parity
\beq \label{R4m}
B_{1}^{(4,-)}=\left[ \matrix{1 & 0 \cr 0 & i}\right],
\quad
B_{2}^{(4,-)}=
\frac{\e^{i\frac{\pi}{4}} }{\sqrt{2}}
\left[ \matrix{\ \    1 & -i \cr -i & \ \ 1}\right],\quad
B_{3}^{(4,-)}=\left[ \matrix{i & 0 \cr 0 & 1}\right]
\eeq
which has not been obtained before in the wave-function approach.

The main idea is to employ the realization of the multi-anyon Pfaffian wave functions as CFT correlation functions
without using their explicit form. The key point is that the precise braid matrices are independent of the distance between 
the particles being braided because they are topological objects. Therefore we could first fuse the particles, which  
we intend to exchange, and then execute braiding by  analytic continuation of the relative coordinate. 
For example, the counterclockwise 
braiding of the quasiparticles with coordinates $\eta_1$  and $\eta_2$ could be executed by the analytic 
continuation along the circle defined by \cite{TQC-NPB,todorov-stanev}
\[
\eta'_1= \frac{\eta_1+\eta_2}{2} + \e^{i\pi t } \frac{\eta_{12}}{2},  \quad
\eta'_2= \frac{\eta_1+\eta_2}{2} - \e^{i\pi t } \frac{\eta_{12}}{2} , \quad
0 \leq t \leq 1.
\]
Thus, if we want to exchange the anyons  with coordinates $\eta_1$  and $\eta_2$  we can first fuse 
$\eta_1 \to \eta_2$ inside the CFT correlator, apply the OPE to extract the short-distance 
singular behavior in terms of the relative coordinate $\eta_{12}=\eta_1-\eta_2$, 
and then execute braiding by a simple permutation $\eta_1 \leftrightarrow \eta_2$  plus analytic 
continuation in $\eta_{12}$, i.e.,  
\[
\eta'_1=\eta_2, \quad \eta'_2=\eta_1,  \quad \left( \textrm{so \ that \ } \eta'_{12} = \e^{i\pi} \eta_{12} \right),
\quad
\eta'_j=\eta_j,  \ \textrm{for} \  j > 2.
\]
This leads to crucial simplifications because the (potentially unknown) CFT correlators after fusion are independent 
of $\eta_{12}$ and their explicit form is not needed since the entire non-analytic behavior comes from 
the short-distance prefactors containing $\eta_{12}$.
%%%%%%%%%%%%%%%%%%%%%%%%%%%%%%%%%%%%%%%%%%%%%%%%%%%%%%%%%%%
\section{Diagonal exchange matrices $B_{1}^{(4,\pm)}$  and $B_{3}^{(4,\pm)}$: fusion rules in the NS sector }
\label{sec:diag}
%%%%%%%%%%%%%%%%%%%%%%%%%%%%%%%%%%%%%%%%%%%%%%%%%%%%%%%%%%%%
When we exchange the quasiholes with coordinates $\eta_1$ and $\eta_2$ or $\eta_3$ and $\eta_4$,
corresponding to the braid generators $B_{1}^{(4,\pm)}$  and $B_{3}^{(4,\pm)}$ respectively,
it is obvious from Eq.~(\ref{4-qh}) that there are even or zero number of $\s$ fields to the right of the
quasiholes being exchanged. Therefore, in order to fuse the quasiholes before braiding we can use the OPE of
two $\s$ fields in the Neveu--Schwarz sector of the Ising model \cite{fst,CFT-book}.

The fusion rules for the $\s_\pm$ fields (\ref{fusion-NS}) lead to the following short-distance OPE in the 
NS sector \cite{fst,CFT-book}
\beq \label{OPE-NS}
\s_\pm(z_1)\s_\pm(z_2)  \mathop{\simeq}_{z_1 \to z_2} z_{12}^{-1/8} \I, \quad  
\s_+(z_1) \s_-(z_2)  \mathop{\simeq}_{z_1 \to z_2} z_{12}^{-1/8} \sqrt{\frac{z_{12}}{2}}\psi(z_2) .
\eeq
However  there is another contribution to the OPE of the quasihole fields in (\ref{fields}) coming from 
the OPE of the  Abelian parts of the quasiholes operators
\beq \label{vertex}
:\e^{i\frac{1}{2\sqrt{2}}\phi(z_1)}: :\e^{i\frac{1}{2\sqrt{2}}\phi(z_2)}:  \mathop{\simeq}_{z_1\to z_2}
z_{12}^{1/8}:\e^{i\frac{1}{\sqrt{2}}\phi(z_2)}: ,
\eeq
which cancels  the factors $z_{12}^{-1/8}$ and from now on we shall skip them
(these factors have been explicitly shown as $\eta_{ab}^{1/8}$ in Eq.~(9) in Ref.~\cite{TQC-NPB}).

It is worth-stressing that the braid matrices $B_1^{(4,\pm)}$ and $B_3^{(4,\pm)}$ must necessarily be diagonal 
\cite{TQC-NPB} because the anyons being exchanged are in the NS sector where the chiral fermion parity is 
preserved \cite{5-2} so no coherent superposition of states with different parity is possible.
Indeed, if we want to exchange $\eta_3$ with $\eta_4$ we  could first fuse them and then do the braiding. However, 
there are no other $\s$ fields to the right of the pair $\s(\eta_3)\s(\eta_4)$ so we have to use the OPE (\ref{OPE-NS})  
in the NS sector and therefore, e.g., the matrix element 
${}_+\la 0|B_3^{(4,+)}|1\ra_+$ must be zero.

In the following subsections we will consider the two cases with positive and negative fermion parity separately.
%%%%%%%%%%%%%%%%%%%%%%%%%%%%%%%%%%%%%%%%%%%%%%%%%%%%%%%%%%%%
\subsection{Positive-parity representation: }
\label{sec:diag+}
%%%%%%%%%%%%%%%%%%%%%%%%%%%%%%%%%%%%%%%%%%%%%%%%%%%%%%%%%%%%
Using the Abelian $\s_\pm$ fields with definite fermion parity and the fusion-path approach \cite{sarma-RMP,clifford} 
to label the anyonic states of matter we can write the computational basis (\ref{01}) for 4 Ising anyons in 
the positive-parity representation as follows \cite{TQC-PRL,TQC-NPB,equiv,clifford}:
\beqa \label{comp-01+}
|0\ra_+ &\equiv & \la \s_+(\eta_1)\s_+(\eta_2)\s_+(\eta_3)\s_+(\eta_4)  \prod_{j=1}^{2N} \psi(z_j) \ra \nn
|1\ra_+  &\equiv & \la  \s_+(\eta_1)\s_-(\eta_2)\s_+(\eta_3)\s_-(\eta_4)  \prod_{j=1}^{2N} \psi(z_j) \ra .  
\eeqa
Recall that quantum information is encoded in the topological charge $\kappa$ of the first pair of $\s$ fields according 
to the rule $|0\ra \leftrightarrow \s_+\s_+ $, $|1\ra \leftrightarrow \s_+\s_- $, while the second pair of $\s$ fields carries no 
information - its purpose is to make the total fermion parity  in (\ref{comp-01+})   trivial, in order for the 
correlation functions to be nonzero, see  Refs.~\cite{TQC-PRL,TQC-NPB,equiv,clifford} for more detail.

To compute the braid matrix $B_1^{(4,+)}$, representing the exchange of the first two anyons, we can first 
fuse $\eta_1 \to \eta_2$  and then implement braiding by $\eta_{12}\to \e^{i\pi} \eta_{12}$. 
The short-distance approximation of the two computational basis states are obtained by 
using the fusion rules (\ref{OPE-NS}) and (\ref{vertex})
\beqa
|0\ra_+  & \mathop{\simeq}_{\eta_1 \to \eta_2} &\la \s_+(\eta_3)\s_+(\eta_4)  \prod_{j=1}^{2N} \psi(z_j) \ra, \nn
|1\ra_+ & \mathop{\simeq}_{\eta_1 \to \eta_2} & \sqrt{\frac{\eta_{12}}{2}}
\la \psi(\eta_2)\s_-(\eta_3)\s_+(\eta_4) \prod_{j=1}^{2N} \psi(z_j) \ra.
\eeqa
Executing the braiding by the analytic continuation $\eta_{12}\to \e^{i\pi} \eta_{12}$ gives 
\beqa
|0\ra_+ &\mathop{\simeq}_{\eta_1 \to \eta_2} &\la \s_+(\eta_3)\s_+(\eta_4)  \prod_{j=1}^{2N} \psi(z_j) \ra  
\mathop{\to}\limits^{B_1} 
\la \s_+(\eta_3)\s_+(\eta_4)  \prod_{j=1}^{2N} \psi(z_j) \ra = |0\ra_+, \nn
|1\ra_+ &\mathop{\simeq}_{\eta_1 \to \eta_2} &  \sqrt{\frac{\eta_{12}}{2}}\la  \psi(\eta_2)\s_+(\eta_3)\s_+(\eta_4)  
\prod_{j=1}^{2N} \psi(z_j) \ra  \mathop{\to}\limits^{B_1} \nn
& \ \to &  \sqrt{\frac{\e^{i\pi}\eta_{12}}{2}}\la  \psi(\eta_2)\s_+(\eta_3)\s_+(\eta_4)  \prod_{j=1}^{2N} \psi(z_j) \ra  = 
i |1\ra_+
\eeqa
so that the first braid generator in the positive-parity representation is simply
\[
B_1^{(4,+)} = \left[\matrix{1 & 0 \cr 0 & i} \right].
\]
Similarly, to compute $B_3^{(4,+)}$ we first fuse $\eta_3 \to \eta_4$, using the fusion rules (\ref{OPE-NS}) 
and (\ref{vertex}) to obtain the short-distance approximation to the computational states,
and then braid $\eta_{34}\to \e^{i\pi} \eta_{34}$  to get, completely in the same way (only replacing $\eta_{12}$ 
with $\eta_{34}$), $B_3^{(4,+)} =B_1^{(4,+)} $.
%%%%%%%%%%%%%%%%%%%%%%%%%%%%%%%%%%%%%%%%%%%%%%%%%%%%%%%%%%%%
\subsection{Negative-parity representation: }
\label{sec:diag-}
%%%%%%%%%%%%%%%%%%%%%%%%%%%%%%%%%%%%%%%%%%%%%%%%%%%%%%%%%%%%
In order to write explicitly the computational basis in the negative-parity representation we could introduce one 
extra Majorana fermion to the right of all $\s$ fields,  still having even number $2N$ of  other Majorana 
fermions. Thus we define the computational basis for 4 Ising anyons in the negative-parity representation
\beqa \label{comp-01-}
|0\ra_-  &\equiv & \la \s_+(\eta_1)\s_+(\eta_2)\s_+(\eta_3)\s_-(\eta_4) \psi(z_0) \prod_{j=1}^{2N} \psi(z_j) \ra \nn
|1\ra_-  &\equiv & \la  \s_+(\eta_1)\s_-(\eta_2)\s_+(\eta_3)\s_+(\eta_4) \psi(z_0) \prod_{j=1}^{2N} \psi(z_j) \ra .  
\eeqa
Again quantum information is encoded in the topological charge $\kappa$ of the first pair of $\s$ fields according to 
$|0\ra \leftrightarrow \s_+\s_+ $, $|1\ra \leftrightarrow \s_+\s_- $, however this time the second pair which fixes the 
total fermion parity of the correlation functions in (\ref{comp-01-}), has opposite parity compared to that in 
(\ref{comp-01+}), see  Refs.~\cite{equiv,clifford} for more detail.

Notice that in general we can insert the extra Majorana fermion between any two pairs of anyons.  
This will define a new basis of computational states in the 
negative-parity representation which is related to (\ref{comp-01-}), in which the extra Majorana fermion is 
to the right of all $\s$ fields,  by a braid transformation that is diagonal, with elements $\pm 1$ on the diagonal
because the Majorana fermion either commutes or anticommutes with any pair of $\s$ fields. 

Now we can compute $B_1^{(4,-)}$ by first fusing $\eta_1 \to \eta_2$ and then taking $\eta_{12}\to \e^{i\pi} \eta_{12}$. 
Because the short-distance expansions of $\s_+(\eta_1) \s_\pm(\eta_2)$  in the NS sector is independent of the 
parity of the other fields in the correlator this braid matrix is the same as for the positive-parity representation, i.e., 
$B_1^{(4,-)} =B_1^{(4,+)}$ (as matrices because they act on different computational bases).

In order to compute $B_3^{(4,-)}$ we first fuse $\eta_3 \to \eta_4$, using the fusion rules (\ref{OPE-NS}) and (\ref{vertex})
to obtain the short-distance approximation to the computational states, which gives
\beqa
|0\ra_-  & \mathop{\simeq}_{\eta_3 \to \eta_4} & \sqrt{\frac{\eta_{34}}{2}} 
\la \s_+(\eta_1)\s_+(\eta_2)  \psi(\eta_4)\psi(z_0)  \prod_{j=1}^{2N} \psi(z_j) \ra, \nn
|1\ra_-  & \mathop{\simeq}_{\eta_3 \to \eta_4} & \la\s_+(\eta_1)\s_-(\eta_2) \psi(z_0)  \prod_{j=1}^{2N} \psi(z_j) \ra.
\eeqa
Executing the braiding $\eta_{34}\to \e^{i\pi} \eta_{34}$ now gives $|0\ra_- \to \e^{i\pi/2}|0\ra_-$ while 
$|1\ra_- \to |1\ra_-$ so that the braid generator for the exchange of the last two $\s$ fields in the 
negative-parity representation is 
\[
B_3^{(4,-)} = \left[\matrix{i & 0 \cr 0 & 1} \right].
\]
%%%%%%%%%%%%%%%%%%%%%%%%%%%%%%%%%%%%%%%%%%%%%%%%%%%%%%%%%%%
\section{Non-diagonal exchange matrices $B_{2}^{(4,\pm)}$: Ising fusion rules in the Ramond sector }
\label{sec:non-diag}
%%%%%%%%%%%%%%%%%%%%%%%%%%%%%%%%%%%%%%%%%%%%%%%%%%%%%%%%%%%%
Just as in the previous sections we are going to use the fact that the braid matrices for coordinate exchanges 
of two anyons are independent  of the distance between them so we can simplify the computation by 
allowing the two anyons to 
fuse, i.e.,  letting $\eta_2 \to \eta_3$ in this case, and reading the exchange phases from the analytic continuation 
of the singular factors containing $\eta_{23}$. 
However, when we exchange the quasiholes with coordinates $\eta_2$ and $\eta_3$ there is one extra $\s$ field
to the right of the quasiholes being exchanged. Therefore, in order to fuse the quasiholes at $\eta_2$ and $\eta_3$ 
before braiding them we need to use the OPE of two $\s$ fields in the Ramond sector of the Ising model 
\cite{fst,CFT-book}.

The OPE of two $\s$ fields in the Ramond sector of the Ising model is more complicated than Eq.~(\ref{OPE-NS})
because the chiral fermion parity in the R-sector is spontaneously broken \cite{5-2} and therefore that OPE might 
contain more terms.
Fortunately this OPE has been explicitly derived in Ref.~\cite{fst}  (see Eq.~(6.47) there, in which we identify 
$\s_e=: \varphi_e$, where $e=\pm$ is the fermion parity) 
from the knowledge of the 4-point function computed in Sect. 6 there and could be written as follows
\beqa \label{OPE-R}
\s_{e_1}(z_1) \s_{e_2}(z_2) |e\ra  &=& \frac{1}{\sqrt{2}z_{12}^{1/8}} 
\left\{ 
\delta_{e_1,e_2} |e\ra +\delta_{e_1,-e_2} |-e\ra \right.\nn
&+&\left. (e.e_2) \sqrt{\frac{z_{12}}{2}} \psi(\sqrt{z_1.z_2}) 
\left( \delta_{e_1,e_2} |-e\ra +\delta_{e_1,-e_2} |e\ra\right)\right\} + \cdots \quad \ \
\eeqa
Recall that in the notation of Ref.~\cite{fst}  the ket-vector $|e\ra$ is defined as the lowest-weight state in the R sector 
with fermion parity $e$, i.e.,  $|e\ra:= \s_e(0)|0\ra$.
We shall use Eq.~(\ref{OPE-R}) in the next Subsections to derive the short-distance approximation of the 
computational state's wave functions in the bases  $\{ |0\ra_+, |1\ra_+ \}$ and $\{ |0\ra_-, |1\ra_- \}$.
%%%%%%%%%%%%%%%%%%%%%%%%%%%%%%%%%%%%%%%%%%%%%%%%%%%%%%%%%%%%
\subsection{ Short-distance OPE  of the computational basis wave functions}
\label{sec:OPE-R}
%%%%%%%%%%%%%%%%%%%%%%%%%%%%%%%%%%%%%%%%%%%%%%%%%%%%%%%%%%%%
In order to simplify the analysis of the fusion process $\eta_2\to \eta_3$ we shall denote the R-sector states
entering the 4-qh wave functions (\ref{comp-01+}) as
\[
|\pm\ra:=\s_{\pm}(\eta_4) \prod_{j=1}^{2N} \psi(z_j) |0\ra ,
\]
and recall that by construction the number $2N$ of Majorana fermions is even.
Let us now apply the R-sector's OPE (\ref{OPE-R}), for $e=+$, to obtain the short-distance expansion of the 
4-anyon computational basis vector $|0\ra_+$ (in the positive-parity representation) defined in Eq.~(\ref{comp-01+}) 
\[
\s_+(\eta_2)\s_+(\eta_3) |+\ra \mathop{\simeq}_{\eta_2\to\eta_3} \frac{1}{\sqrt{2}\eta_{23}^{1/8}}
\left\{  |+\ra + (+.+) \sqrt{\frac{\eta_{23}}{2}} \psi(\sqrt{\eta_2\eta_3}) |-\ra  \right\}.
\]
Then, multiplying from the left by $\la 0| \s_{+}(\eta_1)$ we obtain
\beq
|0\ra_+  \mathop{\simeq}_{\eta_2\to\eta_3}  \frac{1}{\sqrt{2}\eta_{23}^{1/8}}
\left\{  \la 0| \s_{+}(\eta_1)  |+\ra 
+ \sqrt{\frac{\eta_{23}}{2}}     \la 0| \s_{+}(\eta_1)    |-\ra  \right\}.
\eeq
Notice that the overall phase factor $\eta_{23}^{-1/8}$ in the above OPE is exactly canceled by the additional 
inverse factor coming from the OPE of the Abelian part of the Ising anyons, i.e., from Eq.~(\ref{vertex}) for 
$z_1=\eta_2$ and  $z_2=\eta_3$, and we  shall remove it from all expressions below. Thus,
recovering the detailed notation,  we get the 
OPE of the first computational-basis state to be
\beqa \label{comp-0+}
|0\ra_+ & \mathop{\simeq}_{\eta_2\to\eta_3} & \frac{1}{\sqrt{2}}
\left\{  \la 0| \s_{+}(\eta_1)  \s_{+}(\eta_4) \prod_{j=1}^{2N} \psi(z_j) |0\ra  \right.\nn
&+& \left. \sqrt{\frac{\eta_{23}}{2}}    
 \la 0| \s_{+}(\eta_1)     \psi(\sqrt{\eta_2\eta_3}) \s_{-}(\eta_4) \prod_{j=1}^{2N} \psi(z_j) |0\ra  \right\}.  
\eeqa
Similarly, for the computational-basis state $|1\ra_+$ we obtain the short-distance expansion
\beqa\label{comp-1+}
|1\ra_+ & \mathop{\simeq}_{\eta_2\to\eta_3} & \frac{1}{\sqrt{2}}
\left\{  \la 0| \s_{+}(\eta_1)  \s_{+}(\eta_4) \prod_{j=1}^{2N} \psi(z_j) |0\ra  \right.\nn
&-& \left. \sqrt{\frac{\eta_{23}}{2}}     
\la 0| \s_{+}(\eta_1)     \psi(\sqrt{\eta_2\eta_3}) \s_{-}(\eta_4) \prod_{j=1}^{2N} \psi(z_j) |0\ra  \right\}.  
\eeqa
Adding and subtracting Eqs.~(\ref{comp-0+}) and (\ref{comp-1+}) we obtain
\beqa\label{comp-pm}
\frac{ |0\ra_+ + |1\ra_+ }{2} &\mathop{\simeq}_{\eta_2\to\eta_3} &
\frac{1}{\sqrt{2}} \la 0| \s_{+}(\eta_1)  \s_{+}(\eta_4) \prod_{j=1}^{2N} \psi(z_j) |0\ra \nn
\frac{ |0\ra_+ - |1\ra_+ }{2} &\mathop{\simeq}_{\eta_2\to\eta_3}&
\frac{1}{\sqrt{2}}  \sqrt{\frac{\eta_{23}}{2} } \la 0| \s_{+}(\eta_1)    \psi(\sqrt{\eta_2\eta_3})  \s_{-}(\eta_4) 
\prod_{j=1}^{2N} \psi(z_j) |0\ra . \quad \quad
\eeqa
Eq.~(\ref{comp-pm}) will be our starting point for the derivation of the braid matrices in the next subsection
because it expresses the correlation functions on the RHS in terms of the computational basis in the LHS in 
the short-distance limit.
%%%%%%%%%%%%%%%%%%%%%%%%%%%%%%%%%%%%%%%%%%%%%%%%%%%%%%%%%%%%
\subsection{Braiding $\eta_2$ with $\eta_3$ in the positive-parity representation}
\label{sec:B_23+}
%%%%%%%%%%%%%%%%%%%%%%%%%%%%%%%%%%%%%%%%%%%%%%%%%%%%%%%%%%%%
The braiding transformation $B_2^{(4,+)}$ is represented by the coordinate exchange
\[
\eta_2 \to \eta_3,  \quad \eta_3 \to \eta_2, \quad \mathrm{so \  that}\quad \eta_{23} \to \e^{i\pi}\eta_{23}.
\]
Applying the coordinate exchange  over $|0\ra_+$ and making analytic continuation in $\eta_{23}$ we get
\beqa\label{B_2+0}
& & B_2^{(4,+)}|0\ra_+    \mathop{\simeq}_{\eta_2\to\eta_3}  \frac{1}{\sqrt{2}}
\left\{  \la 0| \s_{+}(\eta_1)  \s_{+}(\eta_4) \prod_{j=1}^{2N} \psi(z_j) |0\ra  \right.\nn
&+& \left. \sqrt{\frac{\e^{i\pi}\eta_{23}}{2}}     
\la 0| \s_{+}(\eta_1)     \psi(\sqrt{\eta_2\eta_3}) \s_{-}(\eta_4) \prod_{j=1}^{2N} \psi(z_j) |0\ra  \right\}.  
\eeqa
Now, using $\sqrt{\e^{i\pi}} = i $,  we can substitute the correlation functions appearing in the RHS of 
Eq.~(\ref{B_2+0}) with the expressions in the LHS of  Eq.~(\ref{comp-pm}) to get
 \[
 B_2^{(4,+)}|0\ra_+    \mathop{\simeq}_{\eta_2\to\eta_3}  \frac{|0\ra_+ +|1\ra_+}{2}   +i  \frac{|0\ra_+ -|1\ra_+}{2}   .
  \]
Repeating the same procedure for the computational-basis state  $|1\ra_+$ we obtain from 
Eq.~(\ref{comp-1+}) 
 \[
 B_2^{(4,+)}|1\ra_+   \mathop{\simeq}_{\eta_2\to\eta_3}  
 \frac{|0\ra_+ +|1\ra_+}{2}   -i  \frac{|0\ra_+ -|1\ra_+}{2}  
  \]
so that the braid matrix  in the basis $\{|0\ra_+, |1\ra_+ \}$ is
\beq\label{B_2+}
 B_2^{(4,+)}  = \frac{1}{2} \left[ \matrix{1+i & 1-i \cr 1-i & 1+i}\right] =
\frac{\e^{i\frac{\pi}{4}}}{\sqrt{2}} \left[ \matrix{1 & -i \cr -i & 1}\right] .
\eeq
The braid matrix obtained in Eq.~(\ref{B_2+}) completely coincides with the one obtained in Ref.~\cite{TQC-NPB}
for $| \eta_{12} \eta_{34} / \eta_{13} \eta_{24} |< 1$ , where it was denoted as $R_{23}^{(4)}$.
%%%%%%%%%%%%%%%%%%%%%%%%%%%%%%%%%%%%%%%%%%%%%%%%%%%%%%%%%%%%
\subsection{Braiding $\eta_2$ with $\eta_3$ in the negative-parity representation}
\label{sec:B_23-}
%%%%%%%%%%%%%%%%%%%%%%%%%%%%%%%%%%%%%%%%%%%%%%%%%%%%%%%%%%%%
Again the computational basis in the negative-parity representation is given by Eq.~(\ref{comp-01-})  
where $2N$ is even. Let us now denote 
\[
\s_\pm(\eta_4)  \psi(z_0)\prod_{j=1}^{2N} \psi(z_j) |0\ra =: |\mp\ra
\]
and apply the OPE (\ref{OPE-R}) to obtain the short-distance version of $|0\ra_-$, i.e.,
\[
\s_+(\eta_2) \s_+(\eta_3) |+\ra = \frac{1}{\sqrt{2}} 
\left\{  |+\ra +\sqrt{ \frac{\eta_{23}}{2} }  \psi\left(\sqrt{\eta_2\eta_3}\right) |- \ra \right\},
\]
and similarly, for the computational-basis state $|1\ra_-$ we can use the OPE  (\ref{OPE-R})  in the form
\[
\s_-(\eta_2) \s_+(\eta_3) |-\ra = \frac{1}{\sqrt{2}} 
\left\{  |+\ra - \sqrt{ \frac{\eta_{23}}{2} }  \psi\left(\sqrt{\eta_2\eta_3}\right) |-\ra \right\},
\]
to obtain (after adding and subtracting the results for the two short-distance approximations) 
\beqa\label{comp-2-pm-3}
\frac{ |0\ra_- + |1\ra_- }{2} &\mathop{\simeq}_{\eta_2\to\eta_3} &
\frac{1}{\sqrt{2}} \la 0| \s_{+}(\eta_1)  \s_{+}(\eta_4) \psi(z_0)\prod_{j=1}^{2N} \psi(z_j) |0\ra \nn
\frac{ |0\ra_- -|1\ra_- }{2} &\mathop{\simeq}_{\eta_2\to\eta_3}&
\frac{1}{\sqrt{2}}  \sqrt{\frac{\eta_{23}}{2} } \la 0|  \s_{+}(\eta_1)    \psi(\sqrt{\eta_2\eta_3})  \s_{-}(\eta_4)  \psi(z_0) \prod_{j=1}^{2N} \psi(z_j) |0\ra .
\nonumber
\eeqa
Because the above equation has the same $\eta_{23}$-short-distance structure like Eq.~(\ref{comp-pm}), executing 
the braiding  $\eta_{23} \to \e^{i\pi}\eta_{23}$ produces the same matrix $B_2^{(4,-)}  = B_2^{(4,+)} $ (as matrices).
%\end{enumerate}
Thus we conclude that  $B_2^{(4,-)} $ is indeed given by Eq.~(\ref{R4m}).
%%%%%%%%%%%%%%%%%%%%%%%%%%%%%%%%%%%%%%%%%%%%%%%%%%%%%%%%%%%%
\section{Braid generators for exchanges of $2n+2$ Ising anyons: wave-function approach}
	\label{sec:induct}
%%%%%%%%%%%%%%%%%%%%%%%%%%%%%%%%%%%%%%%%%%%%%%%%%%%%%%%%%%%%
To summarize our results for the exchanges of 4 Ising anyons, we note that  the generators of the negative-parity 
 representation (\ref{R4m}) of $\B_4$ 
completely coincide with those obtained directly from the $\gamma$ matrices for SO$(4)$ in Sect. III of 
Ref.~\cite{equiv}, i.e.,
\[
B_j^{(4,-)} = R_j^{(2,-)}, \quad j=1,2,3 .
\]
In contrast, the second generator of the positive-parity representation of $\B_4$ is different from that obtained in the 
$\gamma$ matrix approach.
Nevertheless, it is easy to see that the positive-parity representation (\ref{R4p}), obtained from the 4-quasiholes 
Pfaffian wave functions, is completely equivalent to the positive-parity representation  in the $\gamma$-matrix approach 
$R_j^{(2,+)}$ (see Eqs. (9) and (10) in Ref.~\cite{equiv}), i.e.,
\[
B_j^{(4,+)} = Z R_j^{(2,+)} Z, \quad j=1,2,3, \quad  Z= \left(B_1^{(4,\pm)}\right)^2 = \left[\matrix{1 & 0 \cr 0 & -1}\right],
\]
and the equivalence transformation is explicitly given by the Pauli matrix $Z$.
This is so because the $Z$ matrix commutes with the diagonal $B_1^{(4,+)}$ and $B_3^{(4,+)}$, however, changes 
the signs of the off-diagonal elements of $B_2^{(4,+)}$. 
It can also be directly seen that the two representations (\ref{R4p})  and (\ref{R4m}) of $\B_4$ are inequivalent, see
 e.g.,  Sect.~III in Ref.~\cite{equiv}. Thus we conclude that the representations (\ref{R4p})  and (\ref{R4m}) of $\B_4$
 are equivalent to the spinor representations of SO$(4)$ with the corresponding parity  \cite{nayak-wilczek,equiv}.

In this section, we shall generalize this result to the braid representations for $2n+2$ Ising anyons.
Our strategy to compute the braid matrices $B_j^{(2n+2,\pm)}$, describing the exchanges of 
$2n+2$ anyons in the Ising representation of the braid group $\B_{2n+2}$, 
would be to fuse some pair of  $\s$ fields, representing one of the qubits in our $n$ qubit system,
which has the effect of projecting out this qubit. The resulting states after fusion will
belong to one of the two representations of $\B_{2n}$ with positive or negative parity
so that we can express the braid matrices $B_j^{(2n+2,\pm)}$ recursively in terms of $B_j^{(2n,\pm)}$. 
More precisely, we shall prove the following recurrence relations:
\\

\noindent
\textbf{Proposition 1:} the $2^n\times 2^n$ dimensional matrices $B_j^{(2n+2,\pm)}$, ($1\leq j \leq 2n+1$)  representing 
the generators of the braid group $\B_{2n+2}$ in the computational bases (\ref{n-qubit-pm}) 
can be expressed recursively in terms of the braid matrices $B_j^{(2n,\pm)}$, ($1\leq j \leq 2n-1$) generating the 
Ising representation of $\B_{2n}$ as follows: 
\begin{enumerate}
\item 
\beq \label{i}
B_j^{(2n+2,+)}=B_j^{(2n+2,-)} \qquad \textrm{for} \quad  1\leq j \leq 2n
\eeq
\item 
\beq \label{ii}
B_j^{(2n+2,\pm)}=B_j^{(2n,\pm)} \otimes \I_2 \qquad \textrm{for} \quad 1\leq j \leq 2n-3
\eeq
\item 
\beq \label{iii}
B_j^{(2n+2,\pm)}=B_{j-2}^{(2n,\pm)} \oplus B_{j-2}^{(2n,\mp)} \quad \textrm{for} \quad 3\leq j \leq 2n+1
\eeq
\end{enumerate}

\noindent
\textbf{Proof:} 
We shall prove this proposition by induction with a base $n=2$. To this end we shall first explicitly prove
statements (i)--(iii) for $n=2$. 
The braid  generators $B_j^{(6,\pm)}$ are $4\times 4$ dimensional matrices defined in the 
computational basis for 6 anyons (corresponding to  2 qubits, encoded in the first two pairs of $\s$ fields, plus one 
extra inert pair, formed by the last two $\s$ fields). For positive parity this basis can be written as \cite{clifford}
\beqa
|00\ra_+ = \la \s_+\s_+\s_+\s_+\s_+\s_+\ra, \quad |01\ra_+ = \la \s_+\s_+\s_+\s_-\s_+\s_-\ra, \nn
|10\ra_+ = \la \s_+\s_-\s_+\s_+\s_+\s_-\ra, \quad |11\ra_+ = \la \s_+\s_-\s_+\s_-\s_+\s_+\ra,
\eeqa
where we skipped for simplicity the product of the even number of Majorana fermions as well as the coordinates $\eta_j$ 
of the fields  $\s_\pm (\eta_j)$. In order to find the braid matrices $B_1^{(6,+)}$  or  $B_2^{(6,+)}$, exchanging
$\eta_1 \leftrightarrow \eta_2$ or $\eta_2 \leftrightarrow \eta_3$, respectively,
we can first fuse $\eta_5\to \eta_6$. The results after fusion are 
computational states from the positive- or negative-parity representations of $\B_4$, i.e.,
\beqa
|00\ra_+ \mathop{\to}\limits_{\eta_5\to \eta_6}  \la \s_+\s_+\s_+\s_+\ra =|0\ra_+ , \quad
|01\ra_+ \mathop{\to}\limits_{\eta_5\to \eta_6}  \la \s_+\s_+\s_+\s_- \psi\ra =|0\ra_-  \nn
|10\ra_+ \mathop{\to}\limits_{\eta_5\to \eta_6}  \la \s_+\s_-\s_+\s_+ \psi \ra =|1\ra_- , \quad
|11\ra_+ \mathop{\to}\limits_{\eta_5\to \eta_6}  \la \s_+\s_-\s_+\s_- \ra =|1\ra_+ .  \nonumber
\eeqa
Let us first compute $B_1^{(6,+)}$. Using the above fusion results for the computational basis,
as well as Eqs.~(\ref{R4p}) and (\ref{R4m}) for $B_1^{(4,\pm)}$, we obtain 
\beqa
B_1^{(6,+)}|00\ra_+ \simeq B_1^{(4,+)}|0\ra_+ = |0\ra_+ \simeq |00\ra_+ \nn
B_1^{(6,+)}|01\ra_+ \simeq B_1^{(4,-)}|0\ra_- = |0\ra_- \simeq |01\ra_+ \nn
B_1^{(6,+)}|10\ra_+ \simeq B_1^{(4,+)}|1\ra_+ = i|1\ra_+ \simeq i|10\ra_+ \nn
B_1^{(6,+)}|11\ra_+ \simeq B_1^{(4,-)}|1\ra_- = i|1\ra_- \simeq i|11\ra_+ , \nonumber
\eeqa
so that $B_1^{(6,+)}=\textrm{diag}(1,1,i,i) = B_1^{(4,+)}\otimes \I_2$.
Next, in the same way we compute $B_2^{(6,+)}$ by using Eqs.~(\ref{R4p}) and (\ref{R4m}) for 
$B_2^{(4,\pm)}$, i.e., we have 
$B_2^{(6,+)}|00\ra_+ \mathop{\to}\limits_{\eta_5\to \eta_6} B_2^{(4,+)}|0\ra_+$ so that
\[
B_2^{(6,+)}|00\ra_+ \simeq \frac{\e^{i\frac{\pi}{4}}}{\sqrt{2}} (|0\ra_+ -i |1\ra_+)
\simeq \frac{\e^{i\frac{\pi}{4}}}{\sqrt{2}} (|00\ra_+ -i |11\ra_+).
\]
Similarly, we have $B_2^{(6,+)}|01\ra_+ \mathop{\to}\limits_{\eta_5\to \eta_6} B_2^{(4,-)}|0\ra_-$ so that
\[
B_2^{(6,+)}|01\ra_+ \simeq \frac{\e^{i\frac{\pi}{4}}}{\sqrt{2}} (|0\ra_- -i |1\ra_-)
\simeq \frac{\e^{i\frac{\pi}{4}}}{\sqrt{2}} (|01\ra_+ -i |10\ra_+).
\]
Continuing in this way  with the states $|10\ra_+$ and  $|11\ra_+$ we find 
\beq \label{B_2}
B_2^{(6,+)}=\frac{\e^{i\frac{\pi}{4}}}{\sqrt{2}} \left[ 
\matrix{1 & 0 & 0 & -i \cr 0 & 1 & -i & 0 \cr  0 & -i & 1 & 0 \cr -i & 0 & 0 & 1 }
\right] .
\eeq
Next, in order to compute the rest of the braid generators we can instead fuse the first two Ising anyons.
This projects out the first qubit so that the computational basis becomes
\beqa
|00\ra_+ \mathop{\to}\limits_{\eta_1\to \eta_2}  \la \s_+\s_+\s_+\s_+\ra =|0\ra_+ , \quad
|01\ra_+ \mathop{\to}\limits_{\eta_1\to \eta_2}  \la \s_+\s_-\s_+\s_- \ra =|1\ra_+  \nn
|10\ra_+ \mathop{\to}\limits_{\eta_1\to \eta_2}  \la \psi \s_+\s_+\s_+\s_-  \ra =|0\ra_- , \quad
|11\ra_+ \mathop{\to}\limits_{\eta_1\to \eta_2}  \la \psi \s_+\s_-\s_+\s_+ \ra =|1\ra_- .  \nonumber
\eeqa
Consider, e.g., the braid matrix $B_3^{(6,+)}$. It is obvious that 
\beqa
B_3^{(6,+)}|00\ra_+\simeq B_1^{(4,+)}|0\ra_+, \quad B_3^{(6,+)}|01\ra_+\simeq B_1^{(4,+)}|1\ra_+, \nn
B_3^{(6,+)}|10\ra_+\simeq B_1^{(4,-)}|0\ra_-, \quad B_3^{(6,+)}|11\ra_+\simeq B_1^{(4,-)}|1\ra_-, \nn
\eeqa
so that $B_3^{(6,+)}=B_1^{(4,+)}\oplus B_1^{(4,-)} = \I_2 \otimes B_1^{(4,+)}$ because $B_1^{(4,-)}=B_1^{(4,+)}$.
Here we used the sign $\oplus$ to denote the direct sum of matrices.
Completely in the same way we find 
$B_4^{(6,+)}=B_2^{(4,+)}\oplus B_2^{(4,-)} = \I_2 \otimes B_2^{(4,+)}$ because $B_2^{(4,-)}=B_2^{(4,+)}$
and $B_5^{(6,+)}=B_3^{(4,+)}\oplus B_3^{(4,-)}=\textrm{diag}(1,i,i,1)$. Notice that the last generator $B_5^{(6,+)}$
is not a tensor product of $\I_2$ and $B_3^{(4,+)}$ because $B_3^{(4,-)}\neq B_3^{(4,+)}$.

Next we have to repeat the above computation of the braid generators $B_j^{(6,-)}$ in the negative-parity representation.
The computational basis is now given by \cite{clifford}
\beqa
|00\ra_- = \la \s_+\s_+\s_+\s_+\s_+\s_-\ra, \quad |01\ra_- = \la \s_+\s_+\s_+\s_-\s_+\s_+\ra, \nn
|10\ra_- = \la \s_+\s_-\s_+\s_+\s_+\s_+\ra, \quad |11\ra_- = \la \s_+\s_-\s_+\s_-\s_+\s_-\ra,
\eeqa
because the total parity of the $\s$ fields must be negative (and this is compensated by the odd number of 
Majorana fermions inside the CFT correlators which are omitted again). It is not difficult to see that the results 
for $B_j^{(6,-)}$  are very similar to  those for 
$B_j^{(6,+)}$ just in each step all $+$ are replaced by $-$ and vice versa. For example 
$B_1^{(6,-)}=\textrm{diag}(1,1,i,i) = B_1^{(4,-)}\otimes \I_2$,  $B_2^{(6,-)}=B_2^{(6,+)}$ (as a consequence of 
$B_2^{(4,-)}=B_2^{(4,+)}$), $B_3^{(6,-)}=B_1^{(4,-)}\oplus B_1^{(4,+)} = \I_2 \otimes B_1^{(4,-)}$ , etc.
The only difference is in the last generator where $B_5^{(6,-)}=B_3^{(4,-)}\oplus B_3^{(4,+)}=\textrm{diag}(i,1,1,i)$
cannot be written as a tensor product. 

The results for the braid matrices $B_j^{(6,\pm)}$ can be summarized as follows: $B_2^{(6,\pm)}$ are equal and 
given by (\ref{B_2}) and the others are given explicitly by
 \beqa
B_1^{(6,\pm)}=B_1^{(4,\pm)}\otimes \I_2,  \quad
B_3^{(6,\pm)}=B_1^{(4,\pm)}\oplus B_1^{(4,\mp)}= \I_2\otimes B_1^{(4,\pm)} , \nn
B_4^{(6,\pm)}= \I_2\otimes B_2^{(4,\pm)} , \quad B_5^{(6,\pm)}= B_3^{(4,\pm)} \oplus B_3^{(4,\mp)} ,
\eeqa
where $B_j^{(4,\pm)}$ are defined in Eqs. (\ref{R4p}) and (\ref{R4m}).
This proves statements (ii) and (iii) for the case $n=2$, which is our induction base. Now it is easy to see that 
in addition these braid matrices satisfy
\[
B_j^{(6,+)}\equiv B_j^{(6,-)} \quad \textrm{for} \quad  1\leq j\leq 4,
\]
which proves the statement (i) for the base $n=2$.

It can also be seen that the positive-parity representation of the braid group $\B_6$ obtained here is completely 
equivalent to the one derived earlier in Refs.~\cite{TQC-PRL,TQC-NPB,universal} and the equivalence is 
established by the braid matrix
$U=B_4^{(6,+)} B_3^{(6,+)} B_5^{(6,+)} B_4^{(6,+)}$ representing the exchange of the pairs 
$(\eta_3,\eta_4)$ and $(\eta_5,\eta_6)$ (recall that in 
the representation of Refs.~\cite{TQC-PRL,TQC-NPB,universal} the inert pair was $\s(\eta_3) \s(\eta_4)$ while 
here the inert pair is 
$\s(\eta_5) \s(\eta_6)$). \\

\noindent
\textit{Induction step:} 
Let us assume that the statements (i)--(iii) are fulfilled for the matrices $B_j^{(2n,\pm)}$.
We must first specify the basis of computational states for $2n+2$ anyons in which the braid matrices are 
represented. The general scheme for representing $n$ qubits in terms of $2n+2$ Ising anyons could be 
described as follows \cite{equiv,clifford}. We group the $2n+2$ fields $\s$ into $n+1$ pairs and encode information 
into the first $n$ pairs: the state of the $i$-th qubit is $|0\ra$ if the $i$-th pair of $\sigma$ fields 
is $\s_+(\eta_{2i-1})\s_+(\eta_{2i})$ (i.e., it fuses to the channel of $\I$) or $|1\ra$ if the $i$-th pair 
is $\s_+(\eta_{2i-1})\s_-(\eta_{2i})$ (i.e., it fuses to the channel of $\psi$). 
The last pair $\s_+(\eta_{2n+1})\s_{c}(\eta_{2n+2})$ contains no information because its state $c$ is determined by 
the requirement  to have a non-zero CFT correlator, i.e., $c=\prod_{i=1}^{2n}c_i$. Thus, the computational states 
in the positive/negative-parity representation of our $n$-qubit system are defined as CFT correlation
 functions of the $(2n+2)$ non-Abelian $\s$ fields and an even/odd number $N$ of Majorana fermions
\beq\label{n-qubit-pm}
|c_1, \ldots, c_n\ra_\pm = \la \s_+\s_{c_1} \cdots \s_+\s_{c_n} \s_+\s_c\prod_{j=1}^N \psi(z_j)\ra.
\eeq
The parity of the representation is denoted by the subscript of the computational basis states: it is `$+$' for positive 
parity (corresponding to even number $N$ of Majorana fermions) and `$-$' for negative parity (corresponding to 
odd number $N$ of Majorana fermions). In other words, $c_j=+$ corresponds to the state $|0\ra$ of the $j$-th qubit, 
while $c_j=-$ corresponds to the state $|1\ra$.
Following our strategy we can first fuse the fields $\s(\eta_{2n-1})$ and $\s(\eta_{2n})$ corresponding to the last 
qubit which has the effect of  projecting out the last qubit, i.e.,
\[
|c_1,c_2,\ldots,c_{n-1},c_n \ra_\pm \ \ \mathop{\longrightarrow}\limits_{\eta_{2n-1}\to \eta_{2n}} \ \ 
\left\{ 
\begin{array}{cc} 
|c_1,c_2,\ldots,c_{n-1}\ra_\pm   & \textrm{if} \ c_n=+ \cr
\pm |c_1,c_2,\ldots,c_{n-1}\ra_\mp   & \textrm{if} \ c_n=- 
\end{array}  
\right. .
\] 
(The sign $\pm$ in front of the $(n-1)$-qubit computational state for $c_n=- $ coincides with the eigenvalue of the
braid generator $B_{2n-1}^{(2n,\mp)}$, however, it is unimportant for our purposes and we will skip it below.)
This means that the computational states after projection will be organized in pairs, such as $|\underbrace{0,1,0\ldots,0,1,1}_{n-1}\ra_+$, $|\underbrace{0,1,0\ldots,0,1,1}_{n-1}\ra_-$,
having exactly the same state of the $(n-1)$-qubit system, however with opposite parity,   on which the 
$(2n+2)$-particle exchange $B_j^{(2n+1,\pm)}$ would act by  $B_j^{(2n,+)}$ and $B_j^{(2n,-)}$.
 Given that we do not touch the last 4 anyons, corresponding to the last qubit and the inert pair, and provided
 that the braid matrices acting trivially on the last 4 anyons in the two representations are the same $B_j^{(2n,+)}=B_j^{(2n,-)}$,  ($1\leq j\leq 2n-2$) because of the inductive step (i),  we arrive at Eq.~(\ref{ii}).
 
On the other hand we can fuse instead the first two $\s$ fields corresponding to projecting out the first qubit. 
Then, in the first half of the computational states, containing $\s_+\s_+$ as a first pair, the result of fusion is $\I$ 
so that the remaining CFT correlation function describes a $(n-1)$-qubit computational state with $2n$ anyons 
and the same parity.  In the second half of the computational states, containing $\s_+\s_-$ as a first pair, the 
result of fusion is $\psi$ so that the remaining CFT correlation function describes a $(n-1)$-qubit computational state
 with $2n$ anyons however with the opposite parity compared to the original one, i.e.,
\[
|c_1,c_2,\ldots,c_{n-1},c_n \ra_\pm \ \  \mathop{\longrightarrow}\limits_{\eta_1\to \eta_2}  \ \
\left\{ 
\begin{array}{cc} 
\ \ \ |c_2,\ldots,c_{n-1},c_n\ra_\pm   & \textrm{if} \ c_1=+ \cr
-|c_2,\ldots,c_{n-1},c_n\ra_\mp   & \textrm{if} \ c_1=- 
\end{array}  
\right. .
\] 
The minus sign multiplying the $(n-1)$-qubit state when $c_1=-1$, which is totally unimportant here
because the braid generators which we want to compute act linearly,  comes from the fact that after fusing the first two
anyons in this case we get one Majorana fermion on the left of all remaining $\s$ fields, which we have to move,
according to our convention, all the way to the right of them as it is in the definition (\ref{n-qubit-pm}) of the 
computational states with negative parity. This produces 
one minus sign for each pair of $\s$'s which is in the state $|1\ra$ but the total sign for this move is always `$-$'.
Next, executing  exchanges on the remaining $2n$ anyons, that do not touch the first qubit,  we immediately find the 
recurrence relations (\ref{iii}) (notice the shift  $j \to j-2$ in the indices  of the braid matrices due to renaming of the 
remaining anyons coordinates $\eta'_j=\eta_{j-2}$ for $3\leq j\leq 2n+2$).
The above mentioned extra minus sign does not change anything because the braidings act linearly. To illustrate 
this consider, e.g., 
the action of $B_6^{(8,+)}$ on the states $|000\ra_+$ and $|100\ra_+$, which after projecting the first qubit will go to 
$\pm |00\ra_\pm$. We have for the first state 
$B_6^{(8,+)}|000\ra_+ \simeq  B_4^{(6,+)} \left(|00\ra_+ \right)=\frac{ \e^{i\frac{\pi}{4}}}{\sqrt{2}}
\left(|00\ra_+   -i |01\ra_+\right)  = \frac{ \e^{i\frac{\pi}{4}}}{\sqrt{2}} \left(|000\ra_+   - i |001\ra_+\right) $, 
while for the second one
\beqa
B_6^{(8,+)}|100\ra_+  \ \  \mathop{\longrightarrow}\limits_{\eta_1\to \eta_2}  \ \
B_4^{(6,-)} \left(-|00\ra_- \right) =-\frac{ \e^{i\frac{\pi}{4}}}{\sqrt{2}}
\left(|00\ra_-   -i |01\ra_-\right)  \mathop{\simeq}\limits_{\eta_1\to \eta_2} \nn
-\frac{ \e^{i\frac{\pi}{4}}}{\sqrt{2}}
\left(-|100\ra_+   + i |101\ra_+\right) =\frac{ \e^{i\frac{\pi}{4}}}{\sqrt{2}}
\left(|100\ra_+   - i |101\ra_+\right) . \nonumber
\eeqa

Finally we must prove statement (\ref{i}) for the matrices $B_j^{(2n+2,\pm)}$. Indeed, we have assumed that 
$B_j^{(2n,-)}= B_j^{(2n,+)}$ for $1\leq j \leq 2n-2$, which is the inductive step (i) for $B_j^{(2n,\pm)}$. 
Then  we can first consider the case when $1\leq j \leq 2n-3$ and use (\ref{ii}) to find
\[
B_j^{(2n+2,-)}= B_j^{(2n,-)} \otimes \I_2 = B_j^{(2n,+)} \otimes \I_2 =B_j^{(2n+2,+)}, \
1\leq j \leq 2n-3.
\]
For the rest of the braid matrices we can use (iii) to prove that for $3\leq j \leq 2n$
\[
B_j^{(2n+2,-)}= B_{j-2}^{(2n,-)} \oplus B_{j-2}^{(2n,+)} =B_{j-2}^{(2n,+)} \oplus B_{j-2}^{(2n,-)} =
B_j^{(2n+2,+)},
\]
because  $B_{j'}^{(2n,+)} =B_{j'}^{(2n,+)} $ for  $1\leq j'\leq 2n -2$ where $j'=j-2$. 
This completes the proof of the Proposition. \\

\noindent
\textbf{Corollary:} 
The recurrence relations (i)--(iii) in the Proposition allow for the following explicit representation for most 
of the braid matrices $B_j^{(2n+2,\pm)}$, which might be useful:
\beq \label{B_2j-1}
B_{2j-1}^{(2n+2,\pm)}= \underbrace{\I_2\otimes \cdots \otimes \I_2}_{j-1} \otimes 
\left[ \matrix{1 & 0 \cr 0 & i}\right] \otimes \underbrace{\I_2\otimes \cdots \otimes \I_2}_{n-j} ,\ \ 
\textrm{for} \ \  1\leq j\leq n,
\eeq
\beq \label{B_2j}
B_{2j}^{(2n+2,\pm)}= \underbrace{\I_2\otimes \cdots \otimes \I_2}_{j-1} \otimes 
\frac{\e^{i\frac{\pi}{4}}}{\sqrt{2}}
\left[ \matrix{1 & 0 & 0 & -i \cr 0 & 1 & -i & 0 \cr 0 & -i & 1 & 0 \cr -i & 0 & 0 & 1 }\right]  
\otimes  \underbrace{\I_2\otimes \cdots \otimes \I_2}_{n-j-1} ,
\eeq
for $n\geq 2$ and  $1\leq j\leq n-1$, as well as
\beq \label{B_2n}
B_{2n}^{(2n+2,\pm)}= \underbrace{\I_2\otimes \cdots \otimes \I_2}_{n-1} \otimes 
\frac{\e^{i\frac{\pi}{4}}}{\sqrt{2}}
\left[ \matrix{1 & -i \cr -i & 1  }\right] ,
\eeq
plus only one more (non-trivial) recursive relation for the last (diagonal) generator
\beq \label{B_2n+1}
B_{2n+1}^{(2n+2,\pm)}=B_{2n-1}^{(2n,\pm)}\oplus B_{2n-1}^{(2n,\mp)}.
\eeq
 with a base $B_{3}^{(4,\pm)}$,  given in Eqs.~(\ref{R4p}) and (\ref{R4m}).
 Notice that braid matrices (\ref{B_2n+1}) are different in the two representations with even or odd number of 
Majorana fermions and cannot be expressed as tensor products of braid operators form
 the braid groups, such as $\B_{2n}$ or $\B_4$, for  smaller number of anyons.
 To illustrate the derivation of the explicit formulas (\ref{B_2j-1}), (\ref{B_2j}) and (\ref{B_2n}) 
consider for example the matrix $B_{4}^{(2n+2,\pm)}$ (i.e., $j=2$ in Eq.~(\ref{B_2j})): on 
the one hand we have from Eq.~(\ref{ii}) 
 \[
 B_{4}^{(2n+2,\pm)}=B_{4}^{(2n,\pm)}\otimes \I_2= \cdots = B_{4}^{(8,\pm)}\otimes \I_{2^{n-3}},
 \]
 where we have used that we can add tensor factors of $\I_2$ to the right, reducing at the same time the value of $2n$, 
until $4 \leq 2n-3$, i.e.,
 until $2n \geq 8$, which gives rise to $(2n-8)/2 +1$ factors and  on the other hand, using Eq.~(\ref{iii}), we have
  \[
 B_{4}^{(2n+2,\pm)}= B_{4}^{(8,\pm)}\otimes \I_{2^{n-3}}=
 \I_2\otimes B_{2}^{(6,\pm)}\otimes \I_{2^{n-3}},
 \]
 because $B_{4}^{(8,\pm)}= B_{2}^{(6,\pm)}\oplus B_{2}^{(6,\mp)}= \I_2\otimes B_{2}^{(6,\pm)}$
(note that $B_{2}^{(6,+)}=B_{2}^{(6,-)}$).
Similarly, combining Eqs.~(\ref{iii}) for $j=2n$ and (\ref{i}) we can verify Eq.~(\ref{B_2n})
\[
B_{2n}^{(2n+2,\pm)}=B_{2n-2}^{(2n,\pm)}\oplus B_{2n-2}^{(2n+2,\mp)} = \I_2\otimes B_{2n-2}^{(2n,\pm)}=
\I_{2^{n-1}}\otimes B_{2}^{(4,\pm)}.
\]

Equations (\ref{B_2j-1}), (\ref{B_2j}), (\ref{B_2n}) and (\ref{B_2n+1}) give the most explicit expressions for the 
generators  $B_{j}^{(2n+2,\pm)}$  of the braid group $\B_{2n+2}$ in the two Ising-model representations with 
opposite parity. These equations also allow us to express the braid matrices, representing the exchanges of Ising 
anyons in the multi-anyon Pfaffian wave functions, in terms of the universal $R$ matrix 
for the Ising model \cite{slingerland-bais,universal} (or, equivalently, for the $\widehat{su(2)}_2$ 
Wess--Zumino-Witten model). Note the crucial role of the projectors to states with definite parity leading
to topological entanglement \cite{universal}, i.e., to the fact that not all braid generators are expressible as tensor products 
of braid matrices with smaller dimensions and the unit matrix $\I_2$.\\
%%%%%%%%%%%%%%%%%%%%%%%%%%%%%%%%%%%%%%%%%%%%%%%%%%%%%%%%%%%%
\section{Proof of the Nayak--Wilczek conjecture}
	\label{sec:proof}
%%%%%%%%%%%%%%%%%%%%%%%%%%%%%%%%%%%%%%%%%%%%%%%%%%%%%%%%%%%%
\noindent
\textbf{Proposition 2:}
The representations of the braid group $\B_{2n+2}$ with positive or negative fermion parity,
constructed from the multi-anyon Pfaffian wave functions, are equivalent to the spinor representations, 
with the corresponding parity, of $\B_{2n+2}$  constructed from the spinor representations of SO$(2n+2)$.
In more detail, the generators $B_j^{(2n+2,\pm)} $ of $\B_{2n+2}$ in the wave-function representations can be 
expressed in terms of the generators  $ R_j^{(n+1,\pm)}$ in the SO$(2n+2)$ representations as follows
\beq \label{equiv}
B_j^{(2n+2,\pm)} = \left( C^{(2n+2,\pm)} \right)^{-1} R_j^{(n+1,\pm)} C^{(2n+2,\pm)},  \quad 1 \leq j \leq 2n+1,
\eeq
where the equivalence matrices for the positive/negative parity ($+/-$) are given explicitly by the product of all diagonal 
generators
\beq \label{C_pm}
C^{(2n+2,\pm)}= \prod_{j=1}^{n+1}  R_{2j-1}^{(n+1,\pm)} = \prod_{j=1}^{n+1}  B_{2j-1}^{(2n+2,\pm)} .
\eeq

\noindent
\textbf{Proof:} 
First of all it is easy to see that the diagonal generators in the wave-function and spinor representations of 
$\B_{2n+2}$ coincide, i.e.,
\[
B_{2j-1}^{(2n+2,\pm)}=R_{2j-1}^{(n+1,\pm)} , \quad 1\leq j\leq n+1
\]
Indeed, as we can see  from \cite{equiv}, the diagonal matrices $B_{2j-1}^{(2n+2,\pm)} $ with indices 
 $1\leq 2j-1\leq 2n-1$, given explicitly  in Eq.~(\ref{B_2j-1})  above, are completely identical to the diagonal matrices $R_{2j-1}^{(n+1,\pm)} $ for  $1\leq 2j-1\leq 2n-1$, given explicitly in Eq.~(26) in Ref.~\cite{equiv}. In addition,
the last diagonal matrices are equal because they satisfy the same recurrence relations 
\[
B_{2n+1}^{(2n+2,\pm)} =B_{2n-1}^{(2n,\pm)} \oplus B_{2n-1}^{(2n,\mp)} \quad
\textrm{or} \quad
R_{2n+1}^{(n+1,\pm)} =R_{2n-1}^{(n,\pm)} \oplus R_{2n-1}^{(n,\mp)} 
\]
with exactly the same bases, $B_{3}^{(4,\pm)}=R_{3}^{(2,\pm)}$. 
Because the equivalence matrices (\ref{C_pm}) 
are diagonal by construction, the matrices $B_{2j-1}^{(2n+2,\pm)}$ and
$R_{2j-1}^{(n+1,\pm)}$ trivially satisfy Eq.~(\ref{equiv}), and therefore we only need to consider the non-diagonal 
matrices. The non-diagonal matrices in the SO$(2n+2)$ representation 
$R_{2j}^{(n+1,\pm)}=P_\pm^{(n+1)}R_{2j}^{(n+1)}P_\pm^{(n+1)}$ can be expressed 
\cite{equiv} as projections of the unprojected matrices ($\s_1$ and $\s_2$ below denote the Pauli matrices)
\[
R_{2j}^{(n+1)}= \underbrace{\I_2\otimes \cdots \otimes \I_2}_{j-1} \otimes \frac{\e^{i\frac{\pi}{4}}}{\sqrt{2}}
\left(  \I_4 -i \s_2 \otimes \s_2\right) \otimes \underbrace{\I_2\otimes \cdots \otimes \I_2}_{n-j} .
\]
On the other hand, Eqs. (\ref{B_2j}) and (\ref{B_2n}) suggest that 
 the braid generators $B_{2j}^{(2n+2,\pm)}$ can be expressed in a similar way as  projections
$B_{2j}^{(2n+2,\pm)}=P_\pm^{(n+1)}B_{2j}^{(2n+2)}P_\pm^{(n+1)}$ (with the same projectors as for 
$R_{2j}^{(n+1,\pm)}$ given in Ref.~\cite{equiv}), of the unprojected matrices
\beq \label{B_2j-un}
B_{2j}^{(2n+2)}= \underbrace{\I_2\otimes \cdots \otimes \I_2}_{j-1} \otimes \frac{\e^{i\frac{\pi}{4}}}{\sqrt{2}}
\left(  \I_4 -i \s_1 \otimes \s_1\right) \otimes \underbrace{\I_2\otimes \cdots \otimes \I_2}_{n-j} ,
\eeq
i.e., if we define the unprojected matrices $B_{2j}^{(2n+2)}$ as  in Eq.~(\ref{B_2j-un}) and apply the 
projectors $P_\pm^{(n+1)}$ as described in  \cite{equiv} then the projected matrices will completely coincide 
with  $B_{2j}^{(2n+2,\pm)}$  as given in Eqs. (\ref{B_2j}) and (\ref{B_2n}).   
Next, we can directly prove that the unprojected matrices are related by
\beq \label{equiv-unproj}
B_{2j}^{(2n+2)}= \left( C^{(2n+2)} \right)^{-1} R_{2j}^{(n+1)} C^{(2n+2)}, \quad 
2\leq 2j \leq 2n, 
\eeq
where the unprojected conjugation matrix is 
\[
C^{(2n+2)}=\underbrace{S\otimes \cdots \otimes S}_{n+1} = \prod_{j=1}^{n+1}R_{2j-1}^{(n+1)}, \quad 
S=\left[ \matrix{1 & 0 \cr 0 & i }\right].
\]
This is simply because of the identity
\[
S^{-1} \s_2 S = \s_1 \quad \Rightarrow \quad \left(S\otimes S\right)^{-1}	 \left( \s_2\otimes \s_2 \right)
 \left(S\otimes S\right) = \s_1 \otimes \s_1.
\]
Now, projecting both sides of (\ref{equiv-unproj}) with the projectors $P_\pm^{(n+1)}$, taking into account that 
$C^{(2n+2)}P_\pm^{(n+1)}=P_\pm^{(n+1)}C^{(2n+2)}$ and $\left(P_\pm^{(n+1)}\right)^2=P_\pm^{(n+1)}$,
 we obtain Eq.~(\ref{equiv})
where the projected equivalence matrix is equal to the product of all diagonal projected braid matrices
and coincides with Eq.~(\ref{C_pm}),
%\[
%C^{(2n+2,\pm)}=P_\pm^{(n+1)}\left(\prod_{j=1}^{n+1}R_{2j-1}^{(n+1)} \right) P_\pm^{(n+1)} = 
%\prod_{j=1}^{n+1}R_{2j-1}^{(n+1,\pm)},
%\]
which completes the proof of Proposition 2 . 

Propositions 1 and 2  ultimately prove that
the Pfaffian correlation functions with $2n+2$ non-Abelian quasiholes at fixed positions indeed belong to one of the 
two inequivalent representations of the braid group $\B_{2n+2}$ whose generators   $B_j^{(2n+2,\pm)}$  can 
be expressed as $\pi/2$ rotations in terms of the SO$(2n+2)$ $\gamma$-matrices 
and identifies the parity in the spinor representations with the fermion parity in the Ising model
\cite{nayak-wilczek,equiv,clifford}.
%%%%%%%%%%%%%%%%%%%%%%%%%%%%%%%%%%%%%%%%%%%%%%%%%%%%%%%%%%%
\section{Conclusions}
%%%%%%%%%%%%%%%%%%%%%%%%%%%%%%%%%%%%%%%%%%%%%%%%%%%%%%%%%%%
In this paper we have consistently derived the braid matrices representing the exchanges of 
4 non-Abelian Ising anyons in both representations with positive and negative fermion parity. To this end we have 
used the fact that the braid matrices are independent of the distance between the braided particles, as well as
the fusion rules for the Ising anyons in both Neveu--Schwarz and Ramond superselection sectors of the Ising model.
In addition we found recurrence relations for the braid matrices $B_j^{(2n+2,\pm)}$ for the exchanges of $2n+2$
Ising anyons as well as explicit  formulas for most of the braid generators in the representations with both parity.
Finally, we have proven that the braid matrices derived from the multi-anyon Pfaffian wave functions are 
completely equivalent to the braid generators derived in the SO$(2n+2)$ spinor approach \cite{nayak-wilczek,equiv}
and have given explicitly the matrices establishing the equivalence in both representations.

\ack
I thank Ivan Todorov, Lyudmil Hadjiivanov and Andr\'e Ahlbrecht for many useful discussions and Galileo Galilei 
Institute for Theoretical Physics in Firenze, 
Italy for hospitality as well as INFN for partial support. The author has been supported as a Research Fellow by 
the Alexander von Humboldt Foundation and by the BG-NCSR under Contract Nos. F-1406 and DO~02-257.\\
%%%%%%%%%%%%%%%%%%%%%%%%%%

\bibliography{FQHE,Z_k,my,TQC}

%\begin{thebibliography}{99}
%\bibitem{geller-loss} M. Geller and D. Loss, Aharonov--Bohm effect in
%the chral Luttinger liquid, Phys. Rev. {\bf B 56}, 9692 (1997).
%\end{thebibliography}
\end{document}